\documentclass[a4,amsmath, amssymb,pra,reprint]{revtex4-2}
\usepackage{graphicx}
\usepackage{color}
\usepackage{mathrsfs}

\begin{document}
	\title{Mode-locking in a semiconductor photonic bandgap laser}
	\author{Emmanuel Bourgon$^1$ , Sylvain Combri\'{e}$^{2}$ , Alexandre Shen$^1$, Nicolas Vaissière$^1$, Delphine Néel$^1$, Fabien Bretenaker $^3$, Alfredo De Rossi$^{2}$}
	\affiliation{$^1$III-V Lab, a joint lab from Nokia, Thales and CEA,  1 av. A. Fresnel, 91767 Palaiseau, France}
	\affiliation{$^{2}$Thales Research and Technology, 1 av. A. Fresnel, 91767 Palaiseau, France}
	\affiliation{$^3$Université Paris-Saclay, ENS Paris-Saclay, CNRS, CentraleSupélec, LuMIn, Orsay, France}
	
\begin{abstract}
Multimode lasers have a very complex dynamics, as expected when oscillators are nonlinearly coupled. Order emerges when the modes lock together; in this case the coherent superposition of the modes results into a periodic train of pulses or a nearly constant  power output with a linearly chirped frequency, for instance. The first is promoted by a saturable absorber, or an equivalent physical mechanism, while the latter is connected to more subtle conditions, such as the fast dynamics of the gain. Here we consider the case of a multimode semiconductor laser with gain provided by quantum wells but without any saturable absorber. The cavity is designed to have a photonic band-gap and very low dispersion. We show, first in theory, that modes can lock together and generate a variety of waveforms which are not trains of pulses nor chirped continuous power waves. Mode locking is observed in experiments on a III-V/Silicon hybrid laser with the cavity made of a suitably tapered grating. Moreover, we find that the mode-locking beatnote is strongly dependent on the injected current: we reach more than 1 GHz modulation amplitude of the beatnote at a modulation frequency of 50 kHz.
The behaviour of the laser is critically determined by the dispersion,  which can be controlled  by the photonic crystal structure. By scaling up the number of interacting modes, this laser source may offer an effective and extremely flexible way of generating waveforms \textit{\`a la carte}. 
\end{abstract}

\maketitle
When gain is provided to a multimode optical resonator, many spectral lines are generated and compete for the available stimulated emission. In some cases, a single mode prevails; otherwise a rich and potentially chaotic dynamics is observed. This complex behaviour is expected, since a multi-mode laser can be modelled as coupled nonlinear oscillators \cite{Lamb1964}. A major achievement of laser physics is to have understood (and harnessed in uncountable applications) complex phenomena such as the phase locking of several laser modes. Mode locking is promoted by the addition of a saturable absorber inside the cavity: absorption decreases  as field intensity increases; as a result, modes interfere to form short and high intensity pulses, which minimizes losses \cite{Haus1975}. Hence, mode locking is explained by a general variational principle. Much less straightforward is mode locking without a saturable absorber. This was reported in quantum cascade lasers (QCL) \cite{Hugi2012} and single section laser diodes \cite{Day2020,Sterczewski2020}. The optical spectrum reveals several lasing lines, however here, in sharp contrast with mode-locking induced by a saturable absorber, the amplitude of the laser emission is weakly modulated and the frequency is linearly chirped. The literature refers to this case as frequency modulated (FM) combs \cite{Hugi2012}. Necessary conditions for this to happen are: spatial hole burning, which reduces mode competition, and nonlinear mode coupling. The solution of the related dynamical equation \cite{Lamb1964} is extremely complex, even for a few modes and gives very little physical insight. Fortunately, FM combs in QCL obey a variational principle, the maximum emission principle (MEP) \cite{Tang1967}. The MEP expects the phases of the modes to lock together such as to maximize the emitted power \cite{Piccardo2019}. Still, the roles of the Kerr nonlinearity and of the dispersion was investigated numerically \cite{Opacak2019}. Burghoff has shown that the FM combs are governed by a phase potential and locking corresponds to the emergence of a phase soliton \cite{Burghoff2020}, similar to mode locking with a saturable losses or microcavity Kerr solitons.\\
A fairly different case is forced mode-locking through periodic modulation inside the cavity. The theory \cite{Haus1975forced} predicts the existence of several mode-locked modes, i.e. supermodes with the spectral distributions of optical energy following the Gaussian-Hermite functions, and establishes a formal analogy with the quantum-mechanical oscillator. The fundamental supermode is stable under slow gain, relative to the cavity round trip time. Yet, with fast gain higher order supermodes become stable leading to a drastic broadening of the optical comb \cite{Heckelmann2023}, even at telecom wavelengths \cite{Marzban2024}.\\     
Here we consider a laser cavity with a photonic potential that shapes its linear (i.e. "cold cavity") modes to a large extent. As a consequence, each mode of the cavity has very different spatial distribution, in contrast with a Fabry-Perot or ring cavity, where modes have approximately the same distribution. The analogy with the quantum mechanical oscillator is more direct as modes are spatial distributions and eigenvalues correspond to photon energies. We will show that this makes the mean field approximation problematic, hence existing soliton theories cannot be used. Moreover, we will show that the MEP cannot be used either, as this assumes gain to be \textit{fast} enough to follow the changes of the field intensity. In contrast with materials with unipolar carrier transport, as QCL, the gain in more conventional quantum well lasers is bound to the interband recombination which is too slow to follow the field amplitude. 
Therefore, existing theories cannot be applied.\\ 
The "photonic potential"  laser can be built using the concept of photonic crystal \cite{Yablonovitch1987,John1987}. Here, a tapered distributed feedback grating (DFG) generates an effective potential acting on the longitudinal localisation of the field, Fig. \ref{fig:cavity_theory}(a). Gain is added  through hybrid integration of the active material \cite{Santis2014}.  We have suggested that such a geometry, through a suitable tapering of the DFG strength, supports modes with spatial envelopes described by the Gauss-Hermite functions and, more importantly, with equi-spaced eigenfrequencies, and predicted mode locking if a saturable absorber is included \cite{Sun2019}.\\ 
Based on a simple rate equation model, accounting for the spatial properties of the modes of this laser cavity, we describe a peculiar form of mode-locking without saturable absorber but capable of generating a variety of temporal waveforms, in contrast with FM combs. Experiments demonstrate mode locking as well as a fast and broadband tuning of the ML beatnote, consistently with our model.
\section*{Theory}
% ============================
%    FIGURE 
\begin{figure}
	\includegraphics[width=1.0\columnwidth]{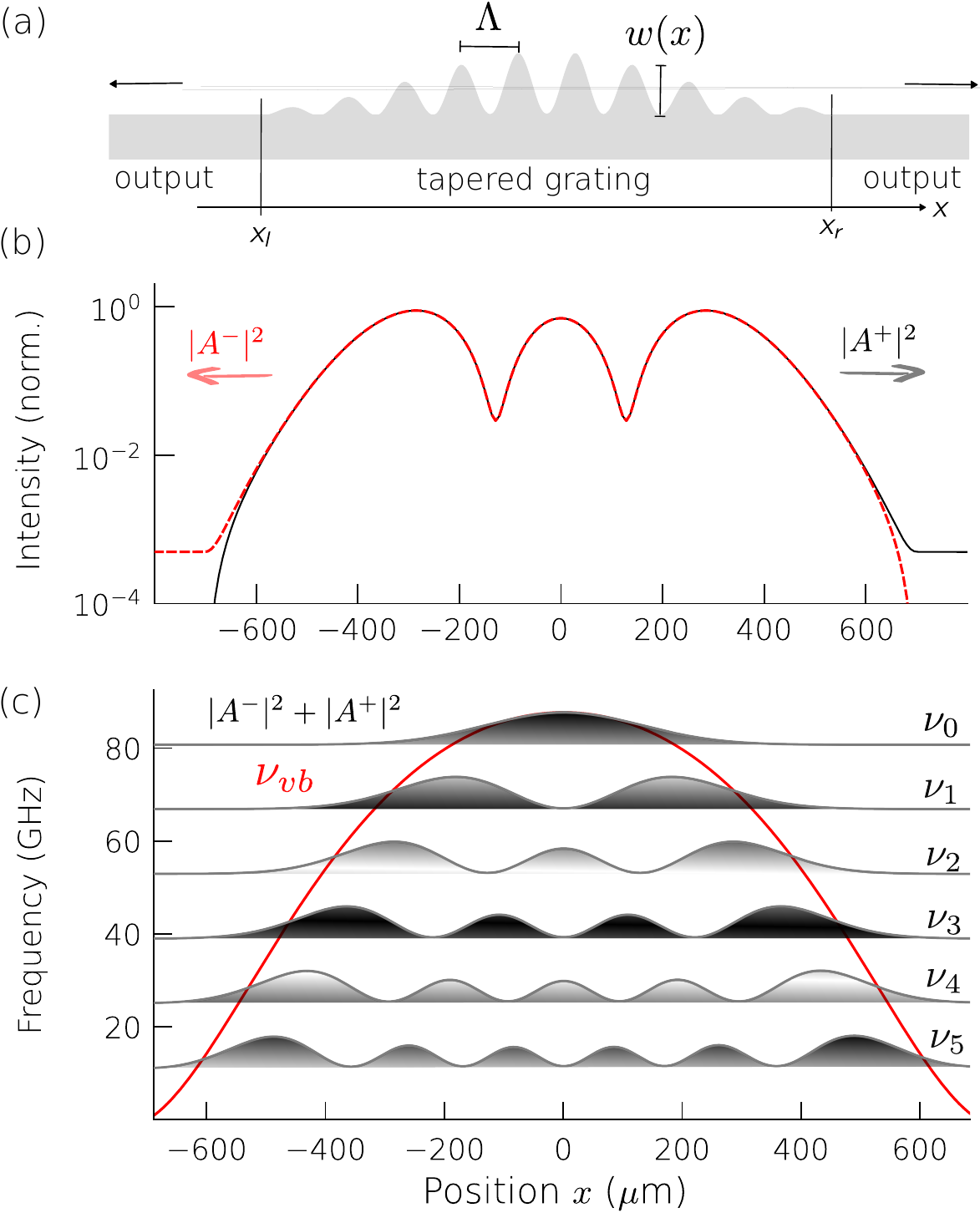}
	\caption{\label{fig:cavity_theory}  (a) Simplified representation of the laser cavity including a tapered DFG on a single-mode waveguide and outputs on both sides; (b) Forward (black) and backward (red) amplitudes $|A^\pm|^2$ describing the $m=2$ mode in a tapered DFG cavity; (c) Eigenmodes of the cavity: normalized intensity $|A^+|^2+A^-|^2$  with vertical offset equal to their frequency $\nu_m$ relative to the surface of the potential well. Note that the latter is formed at the lower edge of the photonic band, therefore the well is oriented "downward" (higher energy correspond to lower optical frequency).}
\end{figure}
% ============================

%  															THEORY: MODEL		
The "photonic potential" and the related existence of a forbidden band gap modifies the density of optical states (DOS) drastically, such that a few, well-confined modes with long lifetime prevail over the weakly localized background in the interaction with the gain material. This is a well known property of photonic crystals \cite{Yablonovitch1987,John1987}. As these modes appear near the edges of spectral transmission gap of the DFG, their linear properties are very well captured by a model involving two linearly coupled, counter-propagating modes with complex amplitudes $A^\pm$. In the experimental setting considered here, the laser cavity is modelled as a tapered DFB made in a waveguide, Fig. \ref{fig:cavity_theory}(a), with  transverse mode $\mathbf{u}(y,z)\exp(\pm\imath k x)$ and amplitudes  $A^\pm(x)$ only depending on the longitudinal coordinate $x$. The total field is represented as the superposition $\mathbf{E}(\mathbf{r},t) =  e^{\imath\omega t} \mathbf{u}(y,z)\left[A^+(x,t)e^{ - \imath k_B x} + A^-(x,t)e^{\imath k_B x}\right] + c.c.$, and $k_B = \pi/\Lambda$ is half the Bragg momentum corresponding to the period $\Lambda$ of the DFG.  
Under the rotating wave approximation, $A^\pm$ is governed by the coupled equations:
\begin{equation}
	\left( \mp\imath v_g\partial_x  +\imath\partial_t \right) A^\pm - \mathcal{K}A^\mp = 0
	\label{eq:linear}
\end{equation}
with the scaled magnitude $\mathcal{K}(x)$ of the modulation of the dielectric permittivity and $v_g$ the group velocity of the waveguide mode. Completed with the Kerr nonlinearity this equation was introduced to describe gap solitons in periodic nonlinear materials \cite{Aceves1989} and more recently in photonic crystals \cite{Malaguti2012}. The longitudinal modes have very large lifetimes (hence large Q factors), as a result of the careful shaping of the coupling strength \cite{Santis2014}. This implies that radiative (i.e. out of plane) leakage is negligible, hence the properties of the modes are perfectly captured by this unidimensional model. Figure \ref{fig:cavity_theory}(b) shows the forward and backward amplitudes for one of the eigenmodes. We note that their profiles are almost identical and highly localized. The only difference is that $A^+$ ($A^-$) is small but non-zero for $x\ge x_r$ ($x\le x_l$), i.e. at the ends of the DFG, where a fraction of the light can escape the cavity. This tiny difference corresponds to very small mirror losses ($<10^{-3}$ in this example), hence large mirror-limited photon lifetime. This is in stark contrast with  typical Fabry-Perot lasers, where mirror losses are large ($>0.1$) and dictate the spatial dependence of the field amplitudes. The angular spectra of the modes are strongly localized around $k = \pm k_B$ \cite{Santis2014}, therefore $A^\pm$ are very smooth functions of $x$.  The fields are normalized such that $S=|A^+|^2 + |A^-|^2$ is the linear density of photons and thus $\int_{x=x_l}^{x_r}Sdx$ is their total number.
Upon proper spatial shaping of the edge of the photonic band, $S$, evaluated on eigenmodes, is distributed as the Hermite-Gauss (HG) functions (if ignoring the interference) and their eigenfrequencies are equispaced, as shown in Fig. \ref{fig:cavity_theory}(c). This suggests an analogy with the quantum mechanical oscillator where the normalized field correspond to the wavefunction and the frequency of the resonance corresponds to the energy of the state \cite{combrie2017}.\\
Let us now consider the implications on the interaction with the semiconductor gain material. The  spatial hole burning (SHB) associated to the interference of the two counter-propagating waves is negligible, because the carrier diffusion length in quantum wells is much larger than the period of the intensity of the standing wave  (equal to $\Lambda$ near the photonic band edge). Therefore, the associated spatial index grating \cite{Agrawal1988, Dong2018} is very weak and can be neglected; on the other hand, diffusion is negligible over the much larger spatial scale of the field $A\pm$.
%
%  															THEORY: FM comb in photonic bandgap lasers		
% ============================
%    FIGURE 
\begin{figure}
	\includegraphics[width=0.9\columnwidth]{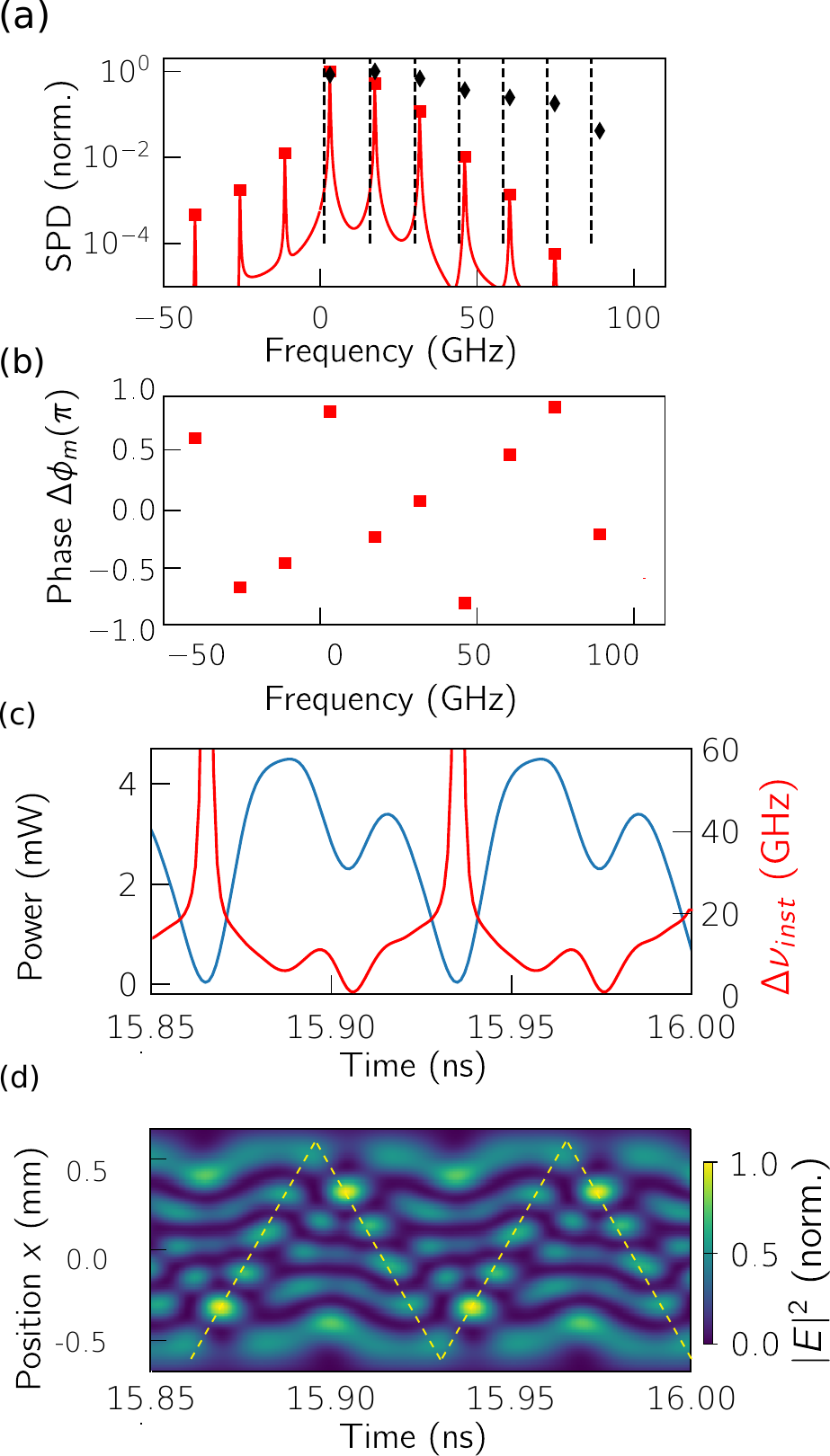}
	\caption{\label{fig:ML_theory}  Locked laser modes. (a) Emitted optical spectral power density (red) and amplitude of the nonlinear modes (makers), inside (black diamonds) and at the output (red squares) compared to the linear modes (dashed); (b) intermodal phases $\Delta\phi_m$ at output; (c) time traces of the output power and nonlinear frequency shift $\Delta\nu_{inst}$, laser started at $t=0$; (d) false-colour time-space maps of the intensity, the dashed line represents a fictive round trip in the cavity with group velocicty $v_g$. The ML beatnote is 14.34 GHz.}
	\end{figure}
	% ============================
	%
Let us point out that the very different spatial distribution of the HG modes promotes a strong SHB which facilitates multimode lasing.
The beating of the modes corresponds to a complex and fast spatio-temporal evolution of the field. This is in conflict with a mean-field theory, which requires the field to evolve slowly within a single round trip in the cavity \cite{Burghoff2020}.\\
The gain is modelled as a homogeneously broadened two-level system \cite{Agrawal1988}, with normalized gain rate $G/G_{max} = 2D(x,t)-1$ (with $D \in [0,1]$) and distributed pump rate $R(x,t)$ (in $s^{-1}$).  
Thus, the evolution of the laser is governed by three coupled nonlinear equations:
\begin{equation}
	\begin{aligned}
		%	\nonumber		
		\left( \mp\imath v_g\partial_x  +\imath\partial_t \right) A^\pm - \mathcal{K}A^\mp &=  \frac{1}{2} (1 +\imath\alpha_H)( G  - \gamma_{ph} )A^\pm \\
		\partial_t D +\gamma_e D  &= R(1-D)  -  \frac{G S}{2N_{l,tr}} 
		%	\label{eq:field}
		\label{eq:rate_eq}
	\end{aligned}
\end{equation}
where $\alpha_H$ is the linewidth broadening factor, $N_{l,tr}$ is the linear density of excited carriers at transparency and $\gamma_{ph}$ its internal losses. \\
The rate equations can be derived from the master equation describing light-matter interaction with a two-level system through the adiabatic elimination of the polarization. This approximation is common in semiconductor laser physics and assumes that intra-band carrier distributions relax to equilibrium very quickly ($< 1$ ps) and therefore, follow adiabatically the  changes in the field \cite{Agrawal1988}. This is justified here by the fact that the spectrum of the interacting modes extends over about 200 GHz and that the spatial structure of HG modes allows nonlinear coupling and linear beating only among nearest neighbours \cite{Sun2019, Sun2020}, with frequency spacing about 10 - 20 GHz. In other words, the impact of spectral hole burning, due to intraband population oscillation, is small compared to interband population oscillations accounted by the rate equation \cite{Agrawal1988}, which justifies the much simpler model introduced here.\\ 
From the numerical solution of eq. \ref{eq:rate_eq} (see Methods) we compute the optical power spectra  at the output shown in Fig. \ref{fig:ML_theory}(a). By filtering around each peak, we retrieve the temporal traces of the phases $\phi_m(t)$ and the amplitudes $A_m(t)$ of the modes $m$, with instantaneous frequency $\nu_m(t)$.  Note that additional peaks appear above the potential well (that is at an offset frequency $<0$) which are due to four wave mixing originating from gain nonlinearity. 
Fig. \ref{fig:ML_theory}(b) shows the intermodal phase \cite{Opacak2019} defined as: $\Delta\phi_m =\phi_{m+1} -\phi_m - \exp[\imath 2\pi (\nu_{m+1}-\nu_m) t]$. Their temporal variance is represented as error bars. In the figure the variance is very small ($<10^{-4}$), therefore all modes are locked and error bars are not visible. The peaks in the output spectrum in Fig. \ref{fig:ML_theory}(a), associated to lower order modes (higher frequencies), are weak, consistently with experiments. This is because of their stronger confinement (hence lower mirror transmission). In contrast, energy is more evenly spread among modes inside the cavity (black markers). \\
In the case of mode-locking associated to (bright) dissipative solitons \cite{Grelu2012} the phases are set such to favour high-energy peaks, while in the case of FM combs these correspond to a linear frequency chirp, such to minimize amplitude modulation \cite{Piccardo2019}. Here, phases are distributed according to a fixed but more complex pattern. We note that this is the case also for the phases of the field inside the cavity. The time-domain analysis of the output in Fig. \ref{fig:ML_theory}(c) reveals modulation with frequency $f_{ML}$, the ML beat frequency, and complex  waveforms in the power, while the corresponding instantaneous frequency $\Delta\nu_{inst}=\frac{1}{2\pi} d\phi/dt$ deviates markedly from a linear chirp, consistently with the non parabolic distribution of the intermodal phases.\\
In contrast to pulsed mode locking, or FM combs, here the intermodal phases take a variety of distributions, which are determined by the choice of the parameters, including the linear dispersion of the cavity 
Interestingly, the coexistence of amplitude and frequency modulated combs was reported recently in quantum dot diode lasers, yet requiring an absorbing section \cite{Hillbrand2020} or more recently through forced mode locking \cite{Heckelmann2023, Marzban2024}. Here the theory establishes a remarkable connection between linear dispersion and the spectral distribution of the supermodes. 
Inside the laser cavity, the spatio-temporal maps of the field intensity, Fig. \ref{fig:ML_theory}(d) reveal an envelope displacing with group velocity estimated to $v_g=2L f_{ML}\approx c/10$, which is about three times smaller than in the hybrid waveguide without grating. This is again a peculiar property of propagation in periodic photonic structures.\\
%
% ============================
	%    FIGURE 
\begin{figure}
	\includegraphics[width=0.9\columnwidth]{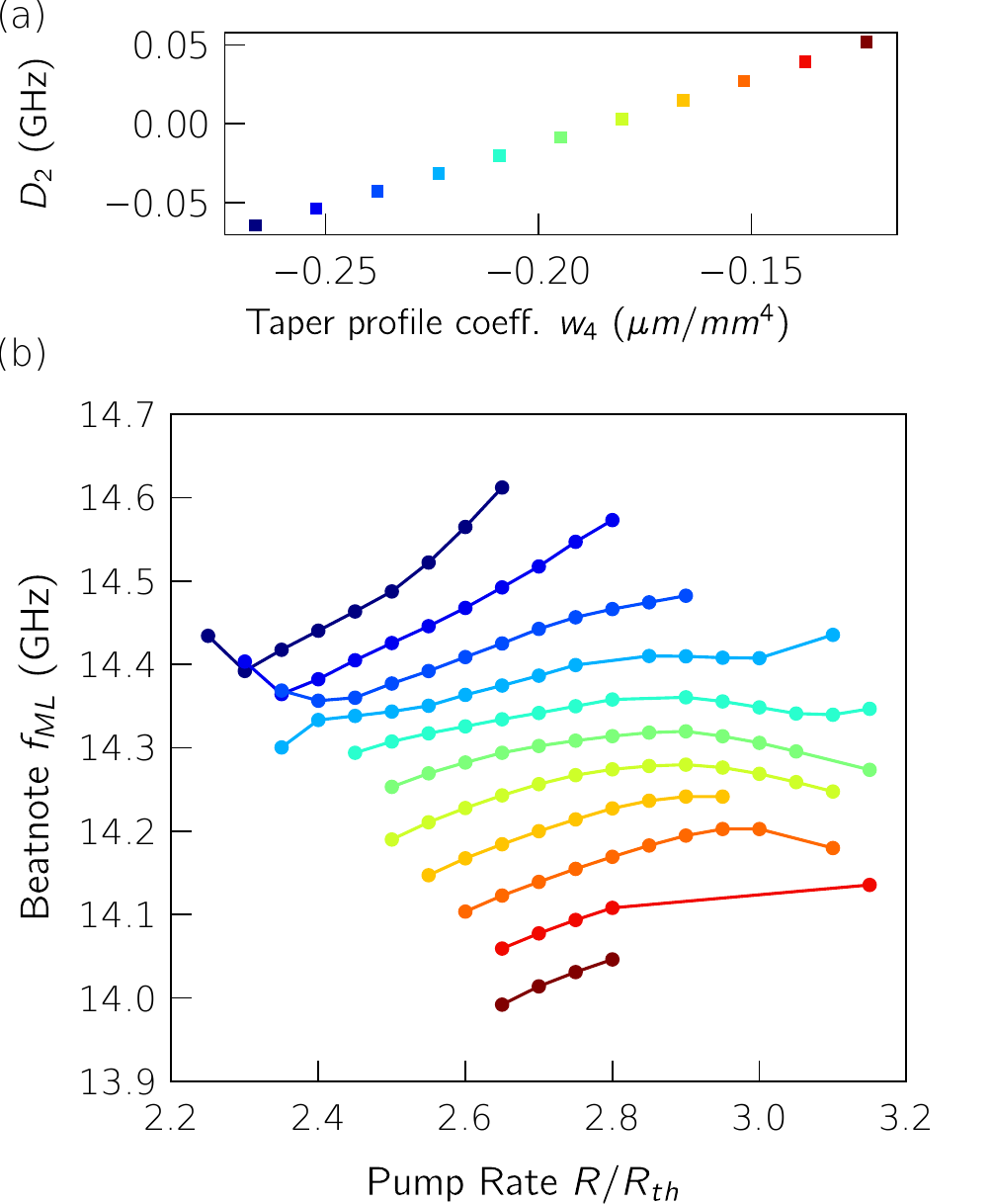}
	\caption{\label{fig:ML_tuning}  (a) Second order dispersion fitting the integrated dispersion $D_{int}(m) = D_2 m^2$ as a function of the parameter controlling the anharmonicity of the photonic band gap $w_4$. (b) ML beat frequency vs the normalized pump rate $R/R_{th}$ as a function of the dispersion (same colour code). The case in Fig. \ref{fig:ML_theory} corresponds to $R/R_{th}=2.5$ and $w_4=-0.215 \mu m/mm^4$.}
\end{figure}
% ============================
In general, modes are locked within a certain range of the pumping rate. This range depends on the parameters, the internal optical losses, the carrier lifetime and the linewidth broadening factor $\alpha_H$. Our modelling predicts model-locking for $\alpha_H>0$, indicating that phase-amplitude coupling is necessary. The parameters chosen are discussed in the Appendix. Most interesting is the dependence on the linear dispersion. To show that, we looked for mode locking regimes as the tapering profile of the grating is modified by changing the the $w_4 x^4$ term in the polynomial representation of the with $w$ of the grating (Fig. \ref{fig:cavity_theory}a). The corresponding second order dispersion is shown in Fig. \ref{fig:ML_tuning}(a); the first order dispersion (the FSR) also increases with $w_4$.   
Fig. \ref{fig:ML_tuning}(b), represents $f_{ML}$ as a function of the  pump rate, as dispersion changes. 
Our model suggests that the beatnote increases not only with the linear FSR, but also with the pump rate.\\
We will now show that this model predicts the salient features observed in the experiments.
\section*{Experiments and discussion}
% ============================
%    FIGURE 
\begin{figure*}
	\includegraphics[width=1.0\textwidth]{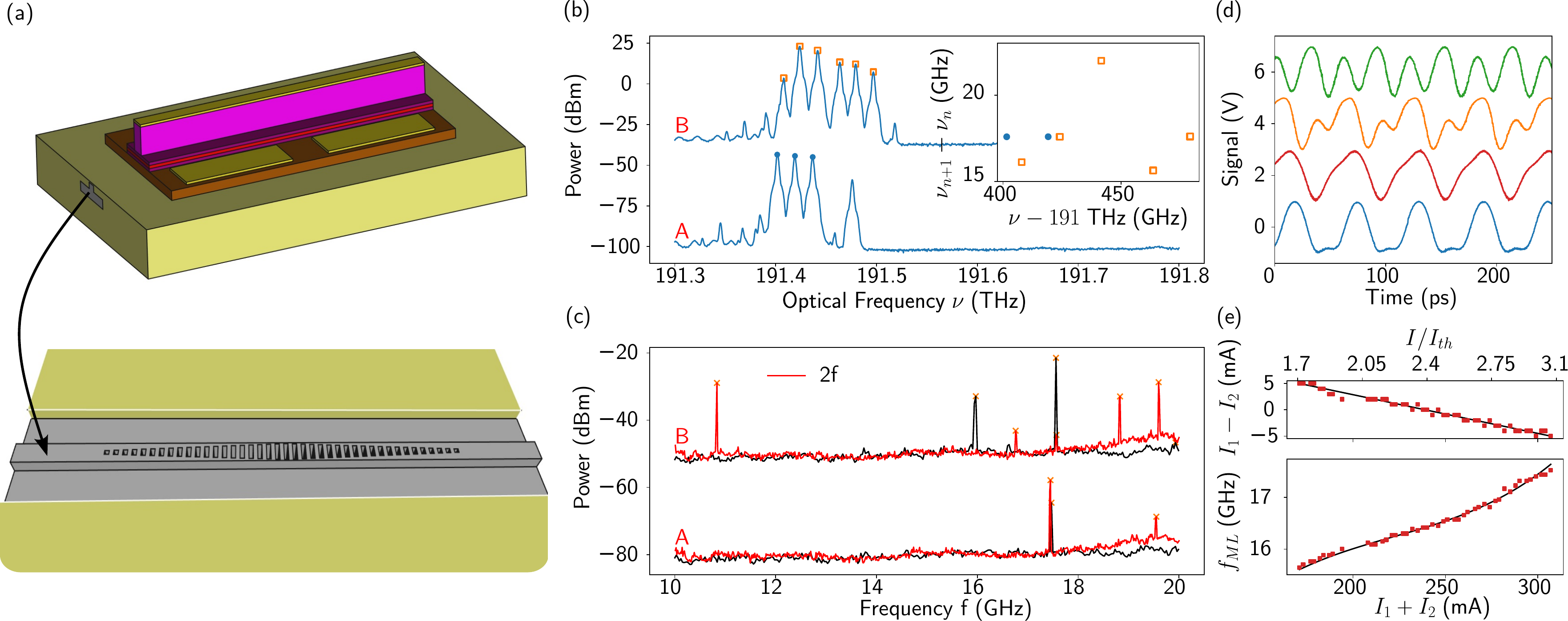}
	\caption{\label{fig:ML_exp}  (a) Artist's representation of III-V (top, purple) on silicon (bottom, gray) laser technology. Silicon is embedded in Silica, serving as optical insulator; (b) optical spectral power density for locked (A) and unlocked (B) modes (vertical offset added) FSR vs mode frequency in the inset; (c) corresponding electric power spectra of the detected optical output [$0-20$GHz] superimposed to the band [$20-40$GHz] with offset (red); (d) temporal traces at the sampling oscilloscope corresponding to different settings of $I_1$, $I_2$ (see Methods); (e) RF beatnote $f_{ML}$ and corresponding current imbalance (top) vs. total injected current (lower axis) and normalized to the total current at threshold (top axis).}
\end{figure*}
\begin{figure}
	\includegraphics[width=1.0\columnwidth]{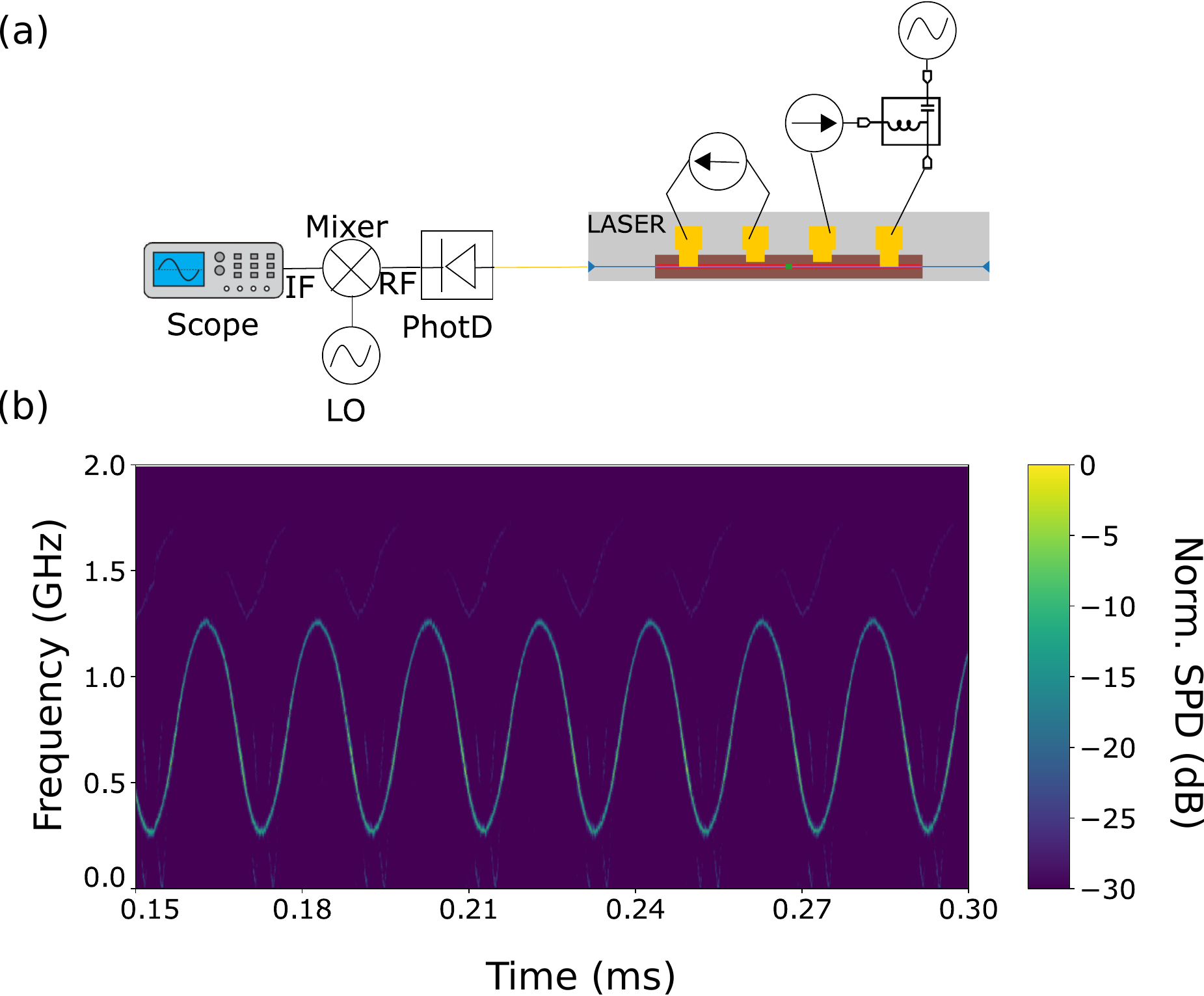}
	\caption{\label{fig:ML_exp_fast}  (a) Setup for fast frequency modulation of the beatnote; (b) spectrogram of the detected signal, after frequency downconversion with a mixer. Here, the laser is modulated at f=50 kHz (see Methods).}
\end{figure}

The laser is based on the hybrid III-V on Silicon technology \cite{duan2014,roelkens2010}, where the III-V material provides gain to a passive, low-loss silicon photonic circuit. Fig. \ref{fig:ML_exp}(a) represents the silicon laser cavity which is a shallow etched waveguide, with the grating shown on the top for the sake of clarity (See methods for details). The silicon layer is bonded to the active material, consisting of a III/V hetero-junction and multiple quantum wells (MQWs). Such DFG lasers, through proper grating tapering, have been shown in experiments to have a large photon lifetime, typically approaching 1 ns. As predicted by the Schawlow-Townes linewidth narrowing theory, this allows a drastic reduction of the phase noise \cite{Santis2014,Santis2018}, compared to uniform DFG lasers.\\ 
The cavity here is larger, with a deeper "photonic potential" and therefore supports several high-Q modes, instead of a single one. As expected from theory, most, if not all, of these modes lase simultaneously, revealing very little mode competition. This is apparent in the optical power spectra, shown in Fig. \ref{fig:ML_exp}(b). Yet, the spectral mode spacing is not exactly regular (squares in the inset). The active region is composed of two electrically separated sections, and therefore independent setting of the bias currents $I_1$,$I_2$. We found areas in the $I_1$,$I_2$ map where the mode spacing becomes strictly identical (case A).
The electrical power spectra of the detected output in Fig. \ref{fig:ML_exp}(c) reveals a single beatnote, perfectly overlapped with its harmonic. In contrast, multiple peaks appear in the unlocked case (B). This demonstrates mode locking. The dependence of the ML condition on the setting of $I_1$ and $I_2$ is consistent with our model predicting ML only for some particular arrangement of linear spectral spacing of the modes. In fact, the two gain sections are both forward biased and the tiny difference between $I_1$ and $I_2$ is causing a slight change in the temperature (about 1 K) and in the carrier density and, consequently, a change in the "photonic potential". As a result the mode spacing is modified.\\
The temporal traces of the detected output are shown in Fig. \ref{fig:ML_exp}(d) for different setting points (values of the pair $I_1$,$I_2$) and reveal trains of pulses with large modulation ratios. This is in stark contrast with the nearly constant amplitude, typical of FM combs (details on the measurement in Methods). Moreover, other waveforms are also observed by changing the operating point of the laser. We note that the waveforms can further be manipulated by controlling their phases with a passive photonic circuit at the output. These two features, again, are consistent with the theory (Fig. \ref{fig:ML_theory}(c)). 
As  predicted by our model (see Fig \ref{fig:ML_tuning}b), the beatnote frequency increases with the pump rate. In fact, Fig. \ref{fig:ML_exp}(e) shows that mode locking is maintained by following a continuous path in the $I_1$,$I_2$ plane and the beatnote $f_{ML}$ changes by about $10\%$ in relative terms. This is larger than expected, but not surprising as our model neglects thermo-refractive effects, which might also change the linear FSR. While a weak dependence of the beatnote frequency on the pump rate was observed long ago in semiconductor lasers \cite{Arahira1997}, here the effect is at least one order of magnitude stronger, for comparable device size. We also note that the range of pump current, relative to the value at the laser threshold, correspond very well to the simulation in Fig. \ref{fig:ML_tuning}(b).\\ 
Fig. \ref{fig:ML_exp}(e) also demonstrates that the beatnote is continuously tuneable all over this range and ML is maintained even when the operating point changes due to a relatively fast ($f_{mod}=50$ kHz) modulation of the current. This is apparent in Fig. \ref{fig:ML_exp_fast}. The detected power signal, shown as a spectrogram in Fig. \ref{fig:ML_exp_fast}(b), reveals large amplitude ($\Delta f\approx 1$ GHz) frequency modulation of the beatnote, which remains well defined  all over the modulation cycle. We note that the product $f_{mod}\Delta f =5\times10^{13}$ Hz$^2$ is comparable to what was achieved with the thin-film Lithium-Niobate technology \cite{YangHe2023}. Fast tuning is consistent with our model, which estimates the time required for reaching locking to few tens of ns to few hundreds of ns. Hence, when the current is modulated with a period of 20 $\mu$s as in Fig. \ref{fig:ML_exp_fast}, the laser can remain locked, leading to the observed fast modulation of the beatnote.\\ 
% ============================
%    FIGURE 
% ============================
%
In summary, these findings allow us to demonstrate the existence of a mode locking regime without saturable absorber (or equivalent mechanism), which is substantially different from the recently investigated  frequency modulated combs. The features of mode locking observed experimentally are captured by a simple distributed rate model, which neglects gain dispersion or fast carrier effects (i.e. spectral hole burning). Yet, fast gain recovery due to stimulated recombination \cite{Dong2018,Marzban2024} is implicit in the rate equations. On the other hand, the peculiar spatial and spectral structure of the modes, existing in a photonic bandgap cavity, has a strong impact on the laser. The photonic bandgap determines the spatial length scales of the fields; this has a strong impact on the carrier dynamics (diffusion, SHB and nonlinear coupling). Therefore, several modes can lase simultaneously because of reduced competition for the gain and may mutually lock owing to four-wave-mixing. Interestingly, this is reminiscent of the Lagrangian description proposed long ago \cite{Schwarz1969}. A necessary condition for locking is dispersion to be small, but not zero, as shown in Fig. \ref{fig:ML_theory}. The beatnote frequency increases strongly with the pump rate, and locking is maintained even if the pump is modulated. Finally, the output is both frequency and amplitude-modulated, with temporal patterns depending on the operating point of the laser (and its dispersion). By adding more modes, in a controlled way, this class of lasers will find applications in microwawe photonics as a flexible pulsed source able to generate controllable temporal waveforms. \\
\section*{Appendix}
\subsection*{III-V on silicon hybrid laser technology}
The laser is a heterogeneously  integrated III-V on silicon photonics laser, fabricated using a standard III-V on Silicon wafer bonding process with a SiO$_{2}$ bonding layer as thin as 20 nm. Optical gain is provided by 6 AlGaInAs quantum wells layers. The laser ridge is buried in BCB. The SOI has been developed by the CEA Leti's 300mm pilot line \cite{Neel2021}. The waveguides are slab waveguides (total thickness: 500 nm. Slab thickness: 265 nm. Ridge thickness: 235 nm). The grating is etched on the backside \cite{Shen2023} of the silicon waveguide, yet this is not critical for the operation of the device. The waveguide  is 2 $\mu$m wide and the slot width is tapered symmetrically between 0.4 $\mu$m and 1.6 $\mu$m. The measurements are performed between 50°C and 60°C to prevent the lasing of another transverse mode at shorter wavelengths ($\sim$1530 nm). This, however has a strong impact on the laser power.
\subsection*{Numerical experiments}
Eqs. \ref{eq:rate_eq}  are solved numerically with the explicit Runge-Kutta fifth order method\footnote{The routine used is \texttt{ParsaniKetchesonDeconinck3S205} } coded in the \texttt{Julia} language \cite{Bezanson2017}. First order spatial derivatives have been discretized using 3rd order finite differences. We used a fixed time step $\Delta t=$0.5 ps and set the size of the spatial mesh to 8 $\mu$m, based on a convergence test. The nonlinear frequencies $\nu_m$ in Fig. \ref{fig:ML_theory}a are computed using the \texttt{esprit} function in the \texttt{DSP.jl} package in \texttt{Julia}, while the phases $\phi_m$ are extracted by filtering the complex amplitude at the output $A^+(x=x_r)$ around each peak $\nu_m$ with a bandwidth of 2 GHz.\\
The locking condition is reached when a) the instantaneous dispersion $|(\nu_{m+1}-\nu_m)/f_{ML}-1| <10^{-4}$ and b) the intermodal phases are stable (their variance is below $0.04\pi$). The markers in figure \ref{fig:ML_tuning}(b) indicate that the conditions are satisfied and the beatnote frequency $f_{ML}$ is unambiguously determined from the spacing of the optical frequencies.  
\subsection*{Parameters and other definitions in the model}
The parameters used in the model are given n in Tab. \ref{tab:param} and characterize the interaction with the gain material, which is governed by the relaxation time of the carriers, the maximum gain rate and the linewidth enhancement (Henry) factor. The carrier damping rate accounts for the internal losses in the optical resonator. These are dominated by Free Carriers Absorption (FCA) in the III-V material and, following \cite{Santis2013}, estimated to $\gamma_{ph}=2\times10^{10} s^{-1}$ The group velocity refers to the III/V hybrid waveguide without the grating. 
The parameters are estimated for a gain section composed of 4 quantum wells (QW) of strained InGaAs (60 $\r{A}$ thick). 
The active region is assumed to be $W=$ 1.8 $\mu m$ wide and to have the same length $L$ as the grating.
The material gain at full inversion is $G_m=$ 5000 cm$^{-1}$, from fig. 2.25 in ref. \cite{coldren-corzine-masha2012}. The maximum gain rate is $G_{max}=G_m v_g \Gamma$, with $\Gamma=0.02$ the overlap between the active material and the field. 
The pump current linear density is $I(x)=2N_{l,tr}R(1-D)$ and the total current is therefore $\int_{x=x_l}^{x_r} I dx$. If $D$ is spatially uniform, its value at lasing threshold is $2D_{ref} = \gamma_{ph}/G_{max}+1$ and the pump rate is $R_{ref} = \gamma_e D_{ref}/(1-D_{ref})$. This value is used to normalize the pump rate. The current injected at threshold is $I_{thr}=2N_{l,tr}R(1-D_{ref})L$. 		
The linear carrier density at transparency  $N_{l,tr}=n_{QW} N_{s,tr} W$, where $N_{s,tr}=10^{12}$cm$^2$  is the sheet carrier density at transparency (per QW), taken from fig. 4.25 in Ref. \cite{coldren-corzine-masha2012}, and $n_{QW}=4$.
\begin{table}
	\begin{tabular}{|l l l|}
		\hline
		\textbf{symbol} &  \textbf{definition} &  \textbf{value}\\
		\hline
		$\gamma_e$&  carrier relaxation rate &  $10^9 s^{-1}$\\
		$\gamma_{ph}$&  photon damping rate & $2\times 10^{10} s^{-1}$ \\
		$v_g$ & group velocity &  $c_0/3.5$\\
		$N_{l,tr}$ & carrier density at tr.&  $8\times10^8$cm$^{-1}$\\
		$G_{max}$&  maximum gain rate&  $10^{12} s^{-1}$\\
		$\alpha_H$&  Henry factor&  3 \\
		\hline
	\end{tabular}
	\caption{parameters used in the model.}
	\label{tab:param}
\end{table}

\subsection*{Measurements}
The device is mounted on a copper support and electrically connected to a dual DC current source (Keithley 2400). The temperature is stabilized at 60 °C in order to align the gain spectrum with the modes of the optical cavity (the DFG design was not well matched here). The laser output is collected with a SMF aligned with a grating coupler.\\ 
Optical spectra are measured with a Yokogawa AQ6370D optical spectrum analyser, the electrical spectra are measured with a 40 GHz (Finisar) photodetector connected to a Rhode \& Schwartz electrical spectrum analyser. Temporal traces are measured using a sampling oscilloscope (Tektronix TDS 8000B)  and a 70 GHz photodiode (Finisar). As the sampling frequency is in the kHz range, the stability of the beatnote frequency is improved by locking the the beatnote $f_{ML}$ to an external frequency synthesizer, through small modulation added to the DC current through a bias tee.\\ 
Spectrograms in Fig. \ref{fig:ML_exp_fast} are measured as follows: a voltage-controlled current source (Koheron DRV400, with 20 mA/V trans-admittance gain) is driven by a sinusoidal voltage signal (HP3245, 1.4 V p-p) and feeds the section $I_2$ (hence $\Delta I_2=$28 mA), while $I_1$ is kept constant. The modulation frequency is set to $f_{mod}=50$ kHz.
The optical output of the laser received by the 40 GHz photodiode is mixed with a RF local oscillator ($f_{LO} = 10\;GHz$ and power $\sim5\; dBm$) and the down-converted signal captured by a 6 GHz (25 GS/s) real-time oscilloscope Tektronix LPD64). The temporal trace is then converted into a spectrogram using the \texttt{scipy.spectrogram} function available in the \texttt{python} language.\\

\bibliography{Bourgon_FMlocking}

\end{document}